\documentclass[twocolumn,aps,amsfonts,amsmath,prd,nofootinbib]
{revtex4-1}
\usepackage[T1]{fontenc}
 \usepackage[latin1]{inputenc}
\usepackage{amssymb}
\usepackage{amsmath}
\usepackage{enumerate}

\newcommand{\N}{{\cal N}}
\def\be{\begin{equation}}
\def\ee{\end{equation}}
\def\ba{\begin{array}}
\def\ea{\end{array}}

\makeatletter

\usepackage{epsf}

\def\be{\begin{equation}}
\def\ee{\end{equation}}
\def\ba{\begin{eqnarray}}
\def\ea{\end{eqnarray}}
\def\beas{\begin{eqnarray*}}
\def\eeas{\end{eqnarray*}}
\def\sla{\raise.15ex\hbox{$/$}\kern-.57em}

\def\E {$E_{7(7)}$}
\def\N {$\cal N$}

\begin{document}

\title{\Large\bf On Absence of  3-loop Divergence in \N=4 Supergravity}

\author{Renata Kallosh}

\

\affiliation{Department of Physics, Stanford University, Stanford, CA 94305, USA}


\begin{abstract}
 We argue that \N=4 supergravity is 3-loop UV finite because the relevant  supersymmetric candidate counterterm is known to be $SL(2, \mathbb{R})\times SO(6)$ invariant, which violates the Noether-Gaillard-Zumino current conservation.  Analogous arguments, based on the universality properties of groups of type E7, also apply to \N=5,6,8 in  4,5,7 loops, respectively, since the  1/\N \, BPS invariants break  duality symmetry between Bianchi identities and quantum corrected vector field equations.

\end{abstract}

\maketitle

\section{Introduction}
A new miraculous cancellation of  the 3-loop ultraviolet divergence was discovered  \cite{Bern:2012cd} at 3-loops  in  d=4 \N=4  supergravity\footnote{The absence of the   3-loop UV divergence in  \N=4 d=4 supergravity was also derived in \cite{Tourkine:2012ip} using the 2-loop heterotic string theory computation and the $R^4$ non-renormalization theorem.  
} using d=4 \N=4  and \N=0 Yang-Mills theory and BCJ color-kinematic duality \cite{Bern:2010ue}.  Pure d=4 \N=4  supergravity  \cite{Cremmer:1977tt} (without vector multiplets) has an 
electro-magnetic  $SL(2, \mathbb{R})\times SO(6)$ duality group ${\cal G}$  which will be central for our discussion of  \N=4.

 Meanwhile, the earlier
recent 
work \cite{Bossard:2011tq} predicted that all \N\, $\geq 4$ supergravities are expected to be UV divergent at  loop order $L=$ \N$-1$, since the  new supersymmetric and duality invariant 1/\N \, BPS  candidate counterterms were constructed at the fully non-linear level.  In particular for \N=4 the 3-loop $R^4$  divergence was predicted and for \N=8 the 7-loop $\partial^8 R^4$  divergence was predicted, complementing the analysis in \cite{Beisert:2010jx}. It is therefore rather important to understand the origin of the cancellation of the UV divergences of 
tens of thousands of high-rank 
tensor integrals in \cite{Bern:2012cd}.

The difference with the previous case of 3-loop UV finiteness of \N=8 \cite{Bern:2007hh} is that the  candidate counterterm \cite{Kallosh:1980fi} was only known at the linear level. But this difference may not be important  since  the duality argument in \cite{Kallosh:2011dp} for explanation of the 3-loop finiteness in \N=8, is also valid for \N=4, as we will show below.  The argument in \cite{Kallosh:2011dp} is based on  duality current conservation and associated with it Noether-Gaillard-Zumino (NGZ)  identity \cite{Gaillard:1981rj}. The argument is valid beyond \N=8 case due to universality property  of  extended supergravity duality groups ${\cal G}$, which belong to  groups  of type E7  \cite{Brown}.

One has to keep in mind that \N=4 supergravity has a 1-loop triangle anomaly \cite{Marcus:1985yy}. Therefore each higher-loop computation may, or may not support  the formal path integral predictions. By looking at Table I in  \cite{Bern:2012cd} it seems likely that  the anomaly may not yet kicked in at the 4-graviton 3-loop level. The role of anomaly requires a separate investigation here. But the underlying path integral prediction \cite{Kallosh:2011dp} for \N=4 supergravity is the $SL(2, \mathbb{R})\times SO(6)$  duality current conservation and associated with it Noether-Gaillard-Zumino (NGZ)  identity \cite{Gaillard:1981rj}.

The old counterterm prediction paradigm was developed in \cite{Kallosh:1980fi}, \cite{Howe:1980th} and applied recently in \cite{Bossard:2011tq}. The new point made in \cite{Kallosh:2011dp} required to  revisit this paradigm: It was shown that the electro-magnetic duality symmetry 
rotating the Bianchi identities $\partial_{\mu} \tilde F^{\mu\nu}=0$ into the vector field equations  $ \partial_{\mu} \tilde G^{\mu\nu}=0$
is always broken when supersymmetric duality invariant quantum corrections are added to classical extended supergravity. This means, quite unexpectedly, that the duality invariant counterterms, including the counterterms constructed in \cite{Bossard:2011tq}, may be forbidden by the requirement of duality invariance of the theory modified by quantum corrections.

A need to revisit the old counterterm paradigm was confirmed in \cite{Bossard:2011ij}. However, it was conjectured there that it is always possible  to restore the duality symmetry in presence of a counterterm, by modifying the original theory. The procedure  of restoration of  duality symmetry of the deformed action was  further developed in \cite{Carrasco:2011jv,Chemissany:2011yv,Broedel:2012gf}. It was demonstrated there that the restoration of  duality broken by the quartic counterterm deformation requires the existence of  the Born-Infeld  type deformation, involving higher derivatives. So far the restoration procedure performed for various  models in \cite{Carrasco:2011jv,Chemissany:2011yv,Broedel:2012gf} was only efficient for $U(1)$ duality models, not including  supergravity.  Moreover, even if a successful Born-Infeld version of \N=4 and \N=8 supergravity were constructed, it is not obvious whether the existence of such new highly nonlinear theories would have any implications for the issue of UV finiteness of the original  \N=4 and \N=8 supergravity, see a discussion of this issue in  \cite{Broedel:2012gf}.

In this paper we will show, along the lines of \cite{Kallosh:2011dp}, that the requirement of duality symmetry  forbids the 3-loop UV divergence in \N=4 supergravity.
In the absence of an alternative explanation of the 3-loop finiteness of \N=4 supergravity, the result of the computation in  \cite{Bern:2012cd} may be viewed as an evidence  that  our duality argument \cite{Kallosh:2011dp} provides a useful tool  for investigation  of UV properties of extended supergravity.

\parskip 9pt

\section{Universality of duality groups of type E7 in extended supergravities}

In all extended supergravities  \N$\geq 4$ scalars are in the coset space ${{\cal G}\over {\cal H}}$ where  the duality group ${\cal G}$ is of  type E7. This includes $SL(2, \mathbb{R})\times SU(4)$, $SU(5,1)$, $SO^*(12)$ and \E\,
for   \N=4,5,6,8  respectively\footnote{For recent studies of the universality in properties of groups of type E7, in application to black holes and cosmology, see \cite{Ferrara:2011dz} and references therein.}. In particular,  duality groups ${\cal G}$ of type E7 in extended supergravity admit a symplectic representation, a doublet $(F, G)$ which transforms in the fundamental
representation of  $Sp(2n, \mathbb{R})$: 
\begin{eqnarray}\label{symplectic}
 \left(
                      \begin{array}{cc}
                  F \\
                 G \\
                      \end{array}
                    \right)'\ =\left(
                      \begin{array}{cc}
                      A&  B \\
                        C & D \\
                      \end{array}
                    \right)  \left(
                      \begin{array}{cc}
                 F \\
               G \\
                      \end{array}
                    \right) \ ,
\end{eqnarray}
whereas the gauge kinetic $n\times n$ matrix  ${\cal N}(\phi)$ transforms via fractional transformation
\be
{\cal N}(\phi)'= (C+D{\cal N})(A+B{\cal N})^{-1} \ .
\label{frac}\ee
Here the vector part of the action is
\be
{\cal L}_v={1\over 4} F \, \cdot \rm{Im} \,  {\cal N} (\phi) \cdot  F + F\,  \, \cdot \rm {Re}  {\cal  N} (\phi)\, \cdot \tilde F \ ,
\label{vecAc}\ee
where the symbol $\cdot$ is used for matrix multiplication.  The scalar part is
\be
{\cal L}_s= {1\over 2} g_{ij}(\phi) \partial _\mu \phi ^i \partial^\mu \phi^j \ ,
\label{S}\ee
where $g_{ij}(\phi)$ is the scalar metric of the nonlinear $\sigma$-model  associated with the ${\cal G}/{\cal H}$ coset space. The dual vector field strength is defined as
\be
\tilde G^{\mu\nu}(F, \phi) \equiv 2 \frac{\delta S_v(F, \phi)}{\delta F_{\mu\nu}} \ .
\label{defG}\ee
The electro-magnetic duality symmetry 
\begin{eqnarray}\label{symplecticdual}
 \left(
                      \begin{array}{cc}
                  \partial_{\mu} \tilde F^{\mu\nu} \\
                  \partial_{\mu} \tilde G^{\mu\nu} \\
                      \end{array}
                    \right)'\ =\left(
                      \begin{array}{cc}
                      A&  B \\
                        C & D \\
                      \end{array}
                    \right)  \left(
                      \begin{array}{cc}
                  \partial_{\mu} \tilde F^{\mu\nu} \\
                  \partial_{\mu} \tilde G^{\mu\nu} \\
                      \end{array}
                    \right) \ ,
\end{eqnarray}
rotating the Bianchi identities $\partial_{\mu} \tilde F^{\mu\nu}=0$ into the vector field equations  $ \partial_{\mu} \tilde G^{\mu\nu}=0$,
is always broken when duality invariant quantum corrections are added to classical extended supergravity. The total quantum corrected action has to transform  under duality \cite{Gaillard:1981rj} as follows:
\be
 {\delta \over \delta F^\Lambda }\Big (  S[F', \varphi']- S[F, \varphi]- {1\over 4} \int (\tilde FCF+ \tilde GBG) \Big)=0 \label{GZ}
 \ee
 Here the duality transformation on vectors acts
so that  the Noether-Gaillard-Zumino  (NGZ) duality current is conserved. The reason for this identity is that $G$ has to transform as in (\ref{symplectic}) but this should also be consistent with its definition given in (\ref{defG})
where the $G$ transformations rules depend on those of $F$ and $\phi$. When the action is deformed, for example by counterterms, so that $S_v= S_v^{cl}+ \lambda S^{ct}$, $G$ is also deformed so that $G(F, \phi)= G^{cl}(F, \phi)+G^{ct}(F, \phi)$. The classical supergravity action satisfy NGZ identity, but the counterterms are duality-invariant, which means that
\be
   S^{ct}[F', \varphi']= S^{ct}[F, \varphi]
\label{inv}\, ,  \ee
which violates the current conservation (\ref{GZ}) for the quantum corrected action, when the counterterms are the only addition to the classical action.

\section{ Counterterm prediction for \N$\geq 4$,  $L=$\N$-1$ UV divergence}

 The true geometric on shell supersymmetric and duality invariant candidate non-BPS counterterms  appear  for the first time in $L=$\N, for example for \N=4 $L=4$, or for \N=8 $L=8$,  \cite{Kallosh:1980fi}, \cite{Howe:1980th}. The status of  1/\N \, BPS invariants, next to geometric ones, was not clear for a very long time. The situation was clarified recently in \cite{Bossard:2011tq} where it was shown that each of these superinvariants can be defined by the integral over the fraction of the superspace, 4(\N -1) fermionic coordinates,  and nevertheless is both supersymmetric as well as duality invariant at the fully non-linear level. These candidate counterterms are given in   \cite{Bossard:2011tq}
\be
I^{\cal N}= \kappa^{2(L-1)}\int d\mu_{({\cal N}, 1. 1)} B_{\alpha\dot \beta}  B^{\alpha\dot \beta}  \ ,
\label{IN}\ee
where
$
L={\cal N}-1$ and $ {\cal N}=4,5,6,8
$.
Here $B_{\alpha\dot \beta}$ is some bilinear combination of the torsion superfield, whose first component is a gaugino field and the measure of integration $d\mu_{({\cal N}, 1. 1)}$ is defined with the help of a harmonic superspace, which allows to single out one particular direction in \N\,  space as a special. For example, in \N=4 
\be
B_{\alpha\dot \beta}\equiv  B_{\alpha\dot \beta l}^{1}\, , \quad 
B_{\alpha\dot \beta k}^{l}\equiv \bar \chi^{lij}_{\dot \beta} \chi_{\alpha kij} \ ,
\ee
where the spinorial superfield $\chi_{\alpha kij}$ and its conjugate $\bar \chi_{\dot \alpha}^{ kij}$ are invariant under the duality group $SL(2, \mathbb{R})\times SO(6)$ and direction 1 (in $i,j,k =1,2,3,4$) is special.

Spinors are invariant under ${\cal G}$-duality, in particular for \N=4 spinorial superfield $\chi_{\alpha kij}$ and its conjugate are $SL(2, \mathbb{R})\times SO(6)$ invariant, for \N=5 they are $SU(5.1)$ invariant, for \N=6 they are $SO^*(12)$ and for \N=8 they are \E\, invariant. This leads to the statement that $I^{{\cal N}= 4}$ is invariant under $SL(2, \mathbb{R})\times SO(6)$,  $I^{{\cal N}= 5}$ is invariant under $SU(5.1)$ 
$I^{{\cal N}= 6}$ is one of the two possible $SO^*(12)$ invariants, and $I^{{\cal N}=8}$ is invariant under \E\, where $I^{\cal N}$ is defined in
(\ref{IN}) for all these cases. Supersymmetry is manifest since the expression is defined in an on shell  superspace. Finally, spinors transform under ${\cal H}$-symmetry, if it is not gauge-fixed, or under the compensating transformation, if it is gauge-fixed, but the counterterms are constructed to be ${\cal H}$-invariant.

It is therefore not accidental that the prediction in \cite{Bossard:2011tq} about the \N=4, L=3 and \N=8, L=7 and intermediate cases, \N=5, L=4 and  N=6, L=5, concerning the universal candidate counterterms in (\ref{IN}) has the flavor of universality for all of these cases. But  it just  turned out \cite{Bern:2012cd} that \N=4, L=3 is free of divergences, whereas the case \N=8, L=7  is beyond our reach,  computationally. 

We will now proceed with the explanation of the argument in \cite{Kallosh:2011dp} which predicts that all these cases are free of divergences. The general case of ${\cal G}$-duality explained in \cite{Gaillard:1981rj}, \cite{Kallosh:2011dp}   for 
\N\,- extended  supergravity and the one for \N=8 with \E\, duality are both complicated technically. The case of \N=4 with $SL(2, \mathbb{R})\times SO(6)$ symmetry of equations of motion and Bianchi identities is, fortunately, relatively simple. 

\section{A toy model of \N=4 supergravity}

We will discuss the \N=4 supergravity  formulation \cite{Cremmer:1977tt} in conventions  of  \cite{Schwarz:1992tn}, which also provides the string theory context of this model. In the toy model we will keep the axion-dilaton and only one vector field, so that only the $SL(2, \mathbb{R})$ duality will be present. The coset space ${{\cal G}\over {\cal H}}$ is  ${SL(2, \mathbb{R})\over U(1)}$.  The scalar part of the action depending on $\tau = \chi+i e^{-\phi}$ is
\be
{\cal L}_s= -\frac{1}{2}\frac{ \partial_{\mu}{\tau}\partial^{\mu}{\bar\tau}}{{\rm Im} \tau ^2}=  (\partial_\mu \phi \partial^\mu \phi + e^{2\phi} \partial _\mu \chi \partial ^\mu \chi) \ .
\label{Ls}\ee
This is a $\sigma$-model action for the ${SL(2, \mathbb{R})\over U(1)}$ coset space, see \cite{Schwarz:1992tn} for details. It is a particular case of the general ${\cal G}/{\cal H}$ scalar action,   given in (\ref{S}).
The action (\ref{Ls}) is $SL(2, \mathbb{R})$ invariant under duality transformation:
\begin{equation}
\tau' = \frac {D\tau + C}{B\tau + A} \, , 
\label{tau}\end{equation}
with real global parameters $A,B,C,D$ restricted by $AD-BC=1$ (in general case in (\ref{symplectic}) each $A,B,C,D$ is given by a  $n\times n$ matrix, restricted by the $Sp(2n, \mathbb{R})$ condition).
The vector part of the bosonic action  is
\begin{equation}
{\cal L}_v=   -{1\over 4  }(e^{-\phi} F^2  + \chi F \tilde  F) \ ,
\label{Lv}\end{equation}
where 
$F_{\mu\nu}=\partial_\mu A_\nu^n -  \partial_\nu A_\mu^n$  and  
$ \tilde F^{\mu\nu}= {1\over 2} e^{-1} \epsilon^{\mu\nu\lambda \sigma} F_{\lambda \sigma}
$.
  Up to a change of conventions between \cite{Schwarz:1992tn} describing \N=4 and generic extended supergravities in \cite{Gaillard:1981rj} the general kinetic term for vectors \N $(\phi)$ can be identified with $\tau$ in \N=4.
 
There is a  Bianchi identity for the  vector field $
\partial^\mu \tilde F_{\mu\nu}=0
$.
To define a duality transformation action on vectors we need to form an $SL(2, \mathbb{R})$ doublet as defined in (\ref{defG})
so that the vector field equations are
$
\partial ^\mu  \tilde G_{\mu\nu}=0
$.
The $SL(2, \mathbb{R})$ symmetry action on the a single vector doublet is given in (\ref{symplectic}) for the $Sp(2n, \mathbb{R})$ with $n=1$.
Under these transformations equations of motion and Bianchi identities are mixed, as shown in (\ref{symplecticdual}).
One can check that the variation of the vector part of the classical vector action  under $SL(2, \mathbb{R})$ transformation of  scalars and vectors given in (\ref{tau}), (\ref{symplectic}),  with $ G$ defined in (\ref{defG}) is
in agreement with the NGZ identity (\ref{GZ}). Note that the action is invariant under ``electric'' transformations with parameters $A, D$ when $B=C=0$. It is only non-invariant when the off-diagonal transformations mixing ``electric'' components with ``magnetic'' , $B, C$ are involved which include a shift of a scalar, $\tau \rightarrow \tau + const$. For example, for $A=1$,  $D=1 $, and $B=\beta$,  $C=\gamma$, 
\be
\delta F=\beta  G\, ,  \qquad \delta G= \gamma F\, ,  \qquad \delta \tau = \gamma- \beta \tau \ .
\ee
That is why the non-trivial part of duality symmetry involves the soft scalar limits, studied in the recent analysis of the supergravity counterterms, for example in \cite{Beisert:2010jx}, but it  also mixes electric and magnetic fields.

\section{Duality invariant counterterms}
  It was important in the proof of duality invariance of $I^{\cal N}$ in (\ref{IN}) that the superfield 
$\chi_{\alpha kij}$ is manifestly invariant under ${\cal G}$ and covariant under ${\cal H}$ for all \N \,$\geq 4$ where scalars are in ${{\cal G}\over {\cal H}}$.  
 In our toy model of \N=4 supergravity with gauge-fixed local ${\cal H}=U(1)$ when the model has only one complex physical scalar $\tau$, an illustration of the point above can be given. Under supersymmetry the first component of the spinor superfield transforms as follows
 \be
 \delta _{\epsilon} \chi_{\alpha ijk}= e^{-\phi/2} F_{\alpha\beta [ij} \epsilon^\beta _{k]} +... \ee
Under global $SL(2, \mathbb{R})$
\be
\Big (e^{-\phi/2}\Big )' = {1\over |B\tau + A|} e^{-\phi/2},\, \quad
(F_{\alpha\beta ij})'= (B\tau + A) F_{\alpha\beta ij}
\ee
Therefore $e^{-\phi/2} F_{\alpha\beta ij}$ transforms with the scalar-dependent phase 
\be
\Big ( e^{-\phi/2} F_{\alpha\beta ij} \Big )' = {B\tau + A\over |B\tau + A|} e^{-\phi/2} F_{\alpha\beta ij} \ ,
\ee
which is a $\tau$-field dependent compensating transformation for local $U(1)$ gauge-fixing. Thus the superfield $\chi_{\alpha ijk}$
also transforms only under the compensating $U(1)$ and the product of two such spinorial superfields 
$ B_{\alpha\dot \beta k}^{l}\equiv \bar \chi^{lij}_{\dot \beta} \chi_{\alpha kij}$
is both $SL(2, \mathbb{R})$ and $U(1)$ invariant.

Thus, if we would look at the bosonic part of the supergravity counterterms, in particular,  $I^{\cal N}$ in (\ref{IN}), we would find that they - being  functions of scalars and vectors - are invariant under  $SL(2, \mathbb{R})$ symmetry  as shown in eq. (\ref{inv}). Therefore the deformed action
\be
S_{def}= S_{cl} + \lambda \, I^{\cal N}
\label{def}\ee
with deformed $SL(2, \mathbb{R})$ doublet $(F, G)$, where
\be
G= G_{cl} + 2 \lambda {\delta I^{\cal N}\over \delta F} \ ,
\ee
does not satisfy the NGZ identity and duality symmetry is broken. In particular, for  the \N=4,  L=3 case the UV divergence $I^{{\cal N}=4}$ would break the duality.

\section{Born-Infeld type  supergravity?}
In \cite{Bossard:2011ij} it was conjectured that it may be possible to develop the deformation of the action (\ref{def}) further, so that the new action
\be
\hat S_{def}= S_{cl} + \lambda S_1 + \lambda^2 S_2+... \lambda^n S_n+...
\label{defn}\ee
is consistent with  NGZ identity (\ref{GZ}), despite the fact that with $S_n=0$ for $n\geq 2$ the duality current conservation is broken. It was suggested in \cite{Bossard:2011ij} that  the duality argument of \cite{Kallosh:2011dp} may not imply UV finiteness in the classes of the models where such construction is possible.

We have studied this proposal in \cite{Carrasco:2011jv}-\cite{Broedel:2012gf} and found that a certain generalization of the procedure or Ref. \cite{Bossard:2011ij} is indeed possible. This lead to the discovery of new, previously unknown models with electro-magnetic duality group ${\cal G}=U(1)$. In particular, the Born-Infeld model with higher derivatives  with initial   deformation of the Maxwell action via open string corrections $\lambda ( \partial F)^4$ with $\lambda = (\alpha')^4$ was completed, a recursive formula for $S_n$ in (\ref{defn}) was found in \cite{Chemissany:2011yv} and all terms of the type $\lambda^n \partial^{4n} F^{2n+2}$  were produced algorithmically. Some large classes of models with non-linear $U(1)$ duality, generalizing the Born-Infeld model with \N=2 global supersymmetry \cite{Kuzenko:2000tg} were constructed in \cite{Broedel:2012gf}.

The reason for the infinite proliferation of Born-Infeld type terms with higher powers of $F$ in extended supergravities is the same as in the original Born-Infeld model \cite{BI}. Once  the Maxwell action is deformed, by quartic in $F$ terms, 
an infinite number of $F^n$ terms has to be added in order to preserve the $U(1)$ duality at the non-linear level. 
 The self-duality property of the Born-Infeld action, 
\be
F\tilde F+ G\tilde G=0 \ ,
\ee 
which is a degenerate case of NGZ identity (\ref{GZ}), was in fact discovered by Schr\"odinger  \cite{Schrodinger} in 1935.

In classical extended supergravities the classical action is universally quadratic in $F$, see eq. (\ref{vecAc}). The 3-loop counterterms $R^4+( \partial F)^4+ R^2 ( \partial F)^2+ (\partial^2 \phi)^4+...$ have terms quartic in $\partial F$, so all higher order terms with more $F$ and more derivatives must be present in (\ref{defn}). When groups of type E7 degenerate to $U(1)$ and extended supergravities degenerates to pure \N=0 Maxwell theory, we know the answer \cite{Chemissany:2011yv} for Born-Infeld model with higher derivatives, satisfying the NGZ constraint at the non-linear level when $G(F)$ depends on all powers of $F$.   It is interesting that the $U(1)$ duality group is a degenerate case of groups of type E7.

The concept of degeneration (when the quartic invariant becomes a perfect square) is easy to illustrate using the \E\,  invariant Cartan-Cremmer-Julia black hole entropy formula \cite{Kallosh:1996uy}, $S= 4\pi  \sqrt J$.  It  depends on one fundamental ${\bf 56}$  $(p^{ij}, q_{ij})$, $i=1,...,8$
\ba
J_{E_{7(7)}}  &=& p^{ij}q_{jk}p^{kl}q_{li}
    -{1\over 4} p^{ij}q_{ij}p^{kl}q_{kl}   \nonumber\\
&+&\, {1\over 96}\epsilon^{ijklmnpq}q_{ij}q_{kl}q_{mn}q_{pq}\nonumber\\
 &+&\, {1\over 96}\epsilon_{ijklmnpq}p^{ij}p^{kl}p^{mn}p^{pq}  \,.
\ea
In \N=4 the symplectic representation is  $\mathbf{ R}=(2,6)$ in $SL(2, \mathbb{R})\times SO(6)$,  and the quartic invariant remains quartic,  not degenerate,  see eqs. (33) in \cite{Ferrara:1996um} 
\ba
J_{SL(2, \mathbb{R})\times SO(6)}= q^2 p^2- (q\cdot p)^2\, .
\ea
Reducing to $U(1)$ with $i=1$ leads to a degeneration of the quartic invariant of groups of type E7
\ba
J_{U(1)}  &=& p^2 q^2 
    -{1\over 4} p^2 q^2= {3\over 4} (pq)^2  
\ea
into a perfect square \cite{Brown,Ferrara:2011dz}. 

From the perspective of the UV finiteness of \N=4 and \N= 8 supergravity, it is important that, at present, the  Born-infeld type duality symmetric model are known only for the subclass of degenerate groups of type E7, namely for $U(1)$ duality models.
This may explain why the duality argument \cite{Kallosh:2011dp}, which was developed for the investigation of the conjectured all-loop finiteness of the  \N=8 supergravity, may also account for the \N=4  case: In both cases the corresponding groups are non-degenerate groups of type E7.

\section{Discussion}
The 3-loop UV finiteness of \N=8  was discovered  \cite{Bern:2007hh} back in 2007.  Five years later, a similar result was obtained in \N=4 supergravity \cite{Bern:2012cd}. 
It is interesting that the origin of miraculous cancellations in both cases may be related to the universality of type E7 duality groups in classical extended supergravities. These dualities (including \E\, and $SL(2, \mathbb{R})\times SO(6)$, respectively) and local extended supersymmetry seem to control the Feynman graphs at the 3-loop quantum level. In \N=8 case other explanations of the 3-loop UV finiteness were proposed over the years, but for \N=4  the  duality current conservation is the only explanation available at present.
More computational data,  especially for  anomaly-free \N=5, L=4 and  \N=6, L=5 will help to test this explanation of the 3-loop \N=4 and \N=8 miracles.
In \N=4 one has to keep in mind that the anomaly may interfere with symmetry expectations starting from L=4. This issue has to be investigated more thoroughly, since it looks plausible that \N=4 L=4 result could be in reach.

In conclusion, we believe that  the duality current conservation argument in  \cite{Kallosh:2011dp}, which explains the just established 3-loop finiteness of \N=4  supergravity \cite{Bern:2012cd},  should be studied more extensively and it may help to clarify the UV properties of extended supergravities.

We are grateful to  Z. Bern, G. Bossard, J. Broedel, J. J. M. Carrasco, W. Chemissany,  M. Duff, D. Freedman,  M. Gunaydin,  A. Linde,  A. Marrani, H. Nicolai, T. Ortin,  R. Roiban and especially S. Ferrara for very stimulating discussions. 
 This work is supported by SITP and NSF grant PHY-0756174.

\end{document}